\documentclass[fleqn,usenatbib]{mnras}
\usepackage{newtxtext,newtxmath}
\usepackage[T1]{fontenc}
\DeclareRobustCommand{\VAN}[3]{#2}
\let\VANthebibliography\thebibliography
\def\thebibliography{\DeclareRobustCommand{\VAN}[3]{##3}\VANthebibliography}
\usepackage{graphicx}	
\usepackage{amsmath}	

\usepackage{amssymb}	
\usepackage{ulem}
\title[Global Properties of Misaligned Galaxies]{SDSS-IV MaNGA: Global Properties of Kinematically Misaligned Galaxies}
\author[Zhou et al.]{Yuren Zhou,$^{1,2,3}$
Yanmei Chen,$^{1,2,3}$\thanks{E-mail: chenym@nju.edu.cn}
Yong Shi,$^{1,2,3}$
Dmitry Bizyaev,$^{4,5}$
Hong Guo,$^{6}$
\newauthor
Min Bao,$^{1,2,3}$
Haitong Xu,$^{1,2,3}$
Xiaoling Yu,$^{1,2,3}$
and Joel R. Brownstein$^{7}$
\\
$^{1}$School of Astronomy and Space Science, Nanjing University, Nanjing 210093, China\\
$^{2}$Key Laboratory of Modern Astronomy and Astrophysics (Nanjing University), Ministry of Education, Nanjing 210093, China\\
$^{3}$Collaborative Innovation Center of Modern Astronomy and Space Exploration, Nanjing 210093, China\\
$^{4}$Apache Point Observatory and New Mexico State University, PO Box 59, Sunspot, NM 88349-0059, USA\\
$^{5}$Sternberg Astronomical Institute, Moscow State University, Moscow 119992, Russia\\
$^{6}$Key Laboratory for Research in Galaxies and Cosmology, Shanghai Astronomical Observatory, Nandan Road 80, Shanghai 200030, China\\
$^{7}$Department of Physics and Astronomy, University of Utah, 115 S. 1400 E., Salt Lake City, UT 84112, USA
}
\date{Accepted XXX. Received YYY; in original form ZZZ}
\pubyear{2020}
\begin{document}
\label{firstpage}
\pagerange{\pageref{firstpage}--\pageref{lastpage}}
\maketitle

\begin{abstract}
We select 456 gas-star kinematically misaligned galaxies from the internal Product Launch-10 of MaNGA survey, including 74 star-forming (SF), 136 green-valley (GV) and 206 quiescent (QS) galaxies. We find that the distributions of difference between gas and star position angles for galaxies have three local peaks at $\sim0^{\circ}$, $90^{\circ}$, $180^{\circ}$. The fraction of misaligned galaxies peaks at $\log(M_*/M_{\odot})\sim10.5$ and declines to both low and high mass end. This fraction decreases monotonically with increasing SFR and sSFR. We compare the global parameters including gas kinematic asymmetry $V_{\mathrm{asym}}$, {\ion{H}{i}} detection rate and mass fraction of molecular gas, effective radius $R_e$, S\'{e}rsic index $n$ as well as spin parameter $\lambda_{R_e}$ between misaligned galaxies and their control samples. We find that the misaligned galaxies have lower {\ion{H}{i}} detection rate and molecular gas mass fraction, smaller size, higher S\'{e}rsic index and lower spin parameters than their control samples. The SF and GV misaligned galaxies are more asymmetric in gas velocity fields than their controls. These observational evidences point to the gas accretion scenario followed by angular momentum redistribution from gas-gas collision, leading to gas inflow and central star formation for the SF and GV misaligned galaxies. We propose three possible origins of the misaligned QS galaxies: (1) external gas accretion; (2) merger; (3) GV misaligned galaxies evolve into QS galaxies.
\end{abstract}


\section{Introduction}
In standard $\Lambda$ cold dark matter ($\Lambda$CDM) cosmological model, galaxies form from primordial density fluctuations and grow through a series of internal and external process. The internal process, or known as the secular evolution, includes stellar winds, supernova explosion, active galactic nucleus (AGN) feedback. The external process includes major/minor mergers and external gas accretion. All these processes are related to gas loss or acquisition, which is the key to regulate the evolution of a galaxy. 

Galaxies can return gas to interstellar space through stellar mass loss and AGN feedback. From conservation law of angular momentum, these returned gas should kinematically aligned with stellar components. Furthermore, if the evolution of galaxies is primarily controlled by the internal process, we would expect kinematic alignment between the stellar and gaseous components. However, about two decades ago, long-slit spectroscopic observations have revealed that phenomenon of gas-star counter-rotation \citep{galletta_detection_1987} is ubiquitous (20$\sim$24\%) in elliptical and lenticular galaxies
\citep{bertola_external_1992, kuijken_search_1996, kannappan_broad_2001}. Due to the development of integral-field spectroscopic (IFS) surveys, on the one hand, they confirm the universality of gas-star 
misalignment (20$\sim$50\%) in early type galaxies \citep{sarzi_sauron_2006, davis_atlas_2011, barrera_tracing_2015, chen_growth_2016, jin_sdss_2016, bryant_sami_2019}. On the other hand, they find a fraction of 2$\sim$5\% gas-star misalignment in blue star forming galaxies \citep{chen_growth_2016, jin_sdss_2016, bryant_sami_2019}.
The misaligned gaseous components in these kinds of galaxies are believed to originate from external processes. They provide an ideal laboratory to investigate the influences of external processes on galaxy evolution. Whether the newly acquired gas triggers/enhances star formation or AGN activities? Whether the structure of the host galaxies is completely reshaped or merely perturbed?

In the last decade, series of work based on the IFS surveys are devoted to study the spatially resolved properties of gas-star kinematically misaligned galaxies, trying to understand their formation mechanisms. Galactic scale gas-star misalignment is believed to originate externally. Galaxy interaction, merger and gas accretion from dwarf companion or cosmic web are all proposed to be its possible origin. Based on $\mathrm{ATLAS}^{\mathrm{3D}}$ survey, \cite{davis_atlas_2011} discover $42\pm5\%$ of early-type galaxies in field showing kinematic misalignments while almost all the galaxies in Virgo cluster have gas-star alignment, which suggest a strong dependence on the environment. \cite{barrera_kinematic_2014, barrera_tracing_2015} compare the $\Delta\mathrm{PA}$ ($\Delta\mathrm{PA}\equiv\left|\mathrm{PA}_{\mathrm{star}}-\mathrm{PA}_{\mathrm{gas}}\right|$, where $\mathrm{PA}_{\mathrm{star}}$ is position angle of stellar velocity field and $\mathrm{PA}_{\mathrm{gas}}$ is PA for gas) distribution between interacting galaxies (different stage of merger) and non-interacting galaxies in CALIFA survey, finding that galaxies with large $\Delta\mathrm{PA}$ ($\gtrsim50^{\circ}$) are dominated by early-type interacting galaxies, compared to small $\Delta{\mathrm{PA}}$ ($<30^{\circ}$) in late-type interacting galaxies and non-interacting control samples, indicating merger is the main formation pathway for kinematic misalignment in early-type galaxies. Combining MaNGA survey with optical images from DESI Legacy survey, \cite{li_impact_2021} find out that the merging remnant fraction is $\sim$10\% higher in quiescent misaligned galaxies with $\Delta\mathrm{PA}>30^{\circ}$ than their co-rotating counterparts, while the fraction is nearly identical between star-forming misaligned galaxies ($\sim$10\%) and co-rotators ($\sim$7\%). Based on the MaNGA survey, the existence of blue counter-rotators has been discovered and they show younger stellar populations with more intense ongoing star formation in the central region \citep{chen_growth_2016, jin_sdss_2016}, suggesting gas accretion in blue star-forming galaxies as an origin of kinematic misalignment. \cite{chen_growth_2016} and \cite{jin_sdss_2016} propose the interaction between pre-existing and accreted gas as an angular momentum loss mechanism. \cite{bryant_sami_2019} discover a similar environment dependence of misaligned fraction using SAMI survey, and point out that mergers are not the primary driver of gas-star misalignment in disc galaxies, while in clusters the misalignment is mainly driven by inner cluster process such as ram pressure stripping. As a summary, the literatures suggest different types of kinematically misaligned galaxies possess different formation mechanisms.

Following the development of observations, kinematically misaligned galaxies are also found and investigated based on cosmological simulations. \cite{taylor_origin_2018} examine 3 gas-star counter-rotating galaxies among 82 galaxies with $10<\log(M_*/M_{\odot})<11.8$ and all of them experience gas accretion from cosmological filaments followed by gas inflow to the galaxy center. \cite{starkenburg_origin_2019} find that $\sim$73\% of counter-rotating galaxies with $9.3<\log(M_*/M_{\odot})<10.7$ experience AGN feedback/flyby through group or cluster before the phenomenon of counter-rotation ($\Delta\mathrm{PA}>90^{\circ}$). \cite{duckworth_decouplingI_2020,  duckworth_decouplingII_2020} find that misaligned galaxies with $\log(M_*/M_{\odot})<10.2$ have enhanced black hole (BH) luminosity and BH growth in IllustrisTNG. Both \cite{starkenburg_origin_2019} and \cite{duckworth_decouplingI_2020,  duckworth_decouplingII_2020} suggest gas removal followed by misaligned gas accretion is a key step to form low mass misaligned galaxies. However, \cite{khoperskov_extreme_2021} discover 25 star-star counter-rotating galaxies with $9.3<\log(M_*/M_{\odot})<10.5$, finding that they all display gas-star misalignments and experience gas accretion but without sudden gas removal by AGN feedback. \cite{bassett_formation_2017} study gas-star counter-rotating S0 galaxies formed through minor merger finding that the key requirement for counter-rotation is the companion galaxy lying in the retrograde orbit with respect to the primary. From IllustrisTNG, \cite{lu_hot_2021} concentrate on the dynamically hot star-forming disc galaxies with $9.7<\log(M_*/M_{\odot})<10.3$ and find out that the counter-rotating galaxies experience less mergers compared to normal disc galaxies. Using Horizon-AGN simulation, \cite{khim_star_2021} estimate relative contribution of different formation mechanisms to the emergence of gas-star misalignment in galaxies with $\log(M_*/M_{\odot})>10$ finding that mergers contribute $\sim$35\%, galaxy interactions (gas accretion from nearby galaxies) contribute $\sim$23\%, environmental interactions such as ram pressure stripping contribute $\sim$21\% and secular evolution (containing smooth gas accretion from filaments) contributes $\sim$21\%. 

In this paper, we select 456 gas-star kinematically misaligned galaxies from the internal Product Launch-10 (MPL-10) in Mapping Nearby Galaxies at Apache Point Observatory \citep[MaNGA,][]{bundy_overview_2015}, 
a new fibre-optic integral field unit (IFU) survey. This is the largest gas-star kinematically misaligned sample so far. As a complementary work of \cite{xu_sdss_2022}, we study the global properties of 
the misaligned galaxies. The paper is organized as follows. In Section~\ref{section:data} we give a short introduction on MaNGA survey and describe our data selection process. In Section~\ref{section:results} the global properties for kinematically misaligned galaxies are studied, including: the distribution of $\Delta\mathrm{PA}$; the fraction of misaligned galaxies as a function of galaxy properties; gas kinematic asymmetry; \ion{H}{i}-detection rate and molecular gas mass; size, morphology, the spin paramters of stars, and the environment. The flat $\Lambda$CDM cosmological model is applied with parameters $H_0=70~\text{km}~\text{s}^{-1}\text{Mpc}^{-1}$, $\Omega_m=0.3$, and $\Omega_{\Lambda}=0.7$ throughout this paper.

\section{data}
\label{section:data}

\subsection{The MaNGA survey}
MaNGA is a core program in the fourth-generation Sloan Digital Sky Survey \cite[SDSS-IV,][]{blanton_sloan_2017} started on July 1, 2014. The program uses the 2.5~m Sloan Foundation Telescope at the Apache Point Observatory \citep{gunn_telescope_2006}. Unlike previous SDSS surveys which obtain single fiber spectra at the centers of target galaxies, MaNGA applies 17 simultaneous hexagonal IFUs \citep{law_observing_2015} with the diameter ranging from $12^{\prime\prime}$ (19 fibers) to $32^{\prime\prime}$ (127 fibers) to explore detailed internal structure of nearby galaxies \citep{drory_manga_2015}, while the mini-bundles with seven fibers are used for flux calibration \citep{yan_sdss_calibration_2016}. It has finished the survey of an unprecedented sample of about 10,000 galaxies in early 2021 with a redshift coverage of $0.01<z<0.15$ \citep{wake_sdss_2017}. The galaxies are selected to span a stellar mass interval of nearly 3 orders of magnitude with a flat distribution in between $9\le\log(M_{*}/M_{\odot})\le11$. The target selection procedure divides galaxies into "Primary" and "Secondary" which extend the radial coverage out to $\sim1.5$ effective radius ($R_e$, Petrosian 50\% light radius) and $\sim2.5R_e$ respectively \citep{yan_sdss_2016}. Two dual-channel BOSS spectrographs provide a wavelength coverage of 3600-10,300{\AA} with a spectral resolution $R\sim2000$ \citep{smee_multi_2013}. Typical 3-hour dithered exposures ensure a per-fiber $r$-band continuum signal-to-noise ratio (S/N) per angstrom of $4-8$ at $1.5R_e$, with much higher S/N towards the center. The $2^{\prime\prime}$ fiber diameter corresponds to a spatial sampling of $1\sim2~\mathrm{kpc}$.

The Data Reduction Pipeline \citep[DRP,][]{law_data_2016} of MaNGA provides sky subtracted and flux calibrated 3D spectra for each galaxy. The Data Analysis Pipeline (DAP) is a survey-led software package for analyzing the spectra produced by the DRP to estimate physical properties.  The DAP \citep{westfall_data_2019} has been developed since 2014 and gone through several versions. In this work, we use the MaNGA sample and DAP products drawn from the MPL-10. The DAP heavily relies on penalized pixel-fitting \citep[pPXF,][]{cappellari_parametric_2004} and a subset of stellar templates from MaSTar library \citep{yan_sdss_2019} to fit the stellar continuum and absorptions in each spaxel. The data products include the measurements of stellar kinematics, nebular emission-line properties (e.g., fluxes, equivalent widths, kinematics), spectral indices of absorption-line (e.g., H$\delta$) and bandhead/color (e.g., TiO, $\mathrm{D}_n$4000).

\subsection{Sample selection}
\begin{figure*}
    \includegraphics[width=2.0\columnwidth]{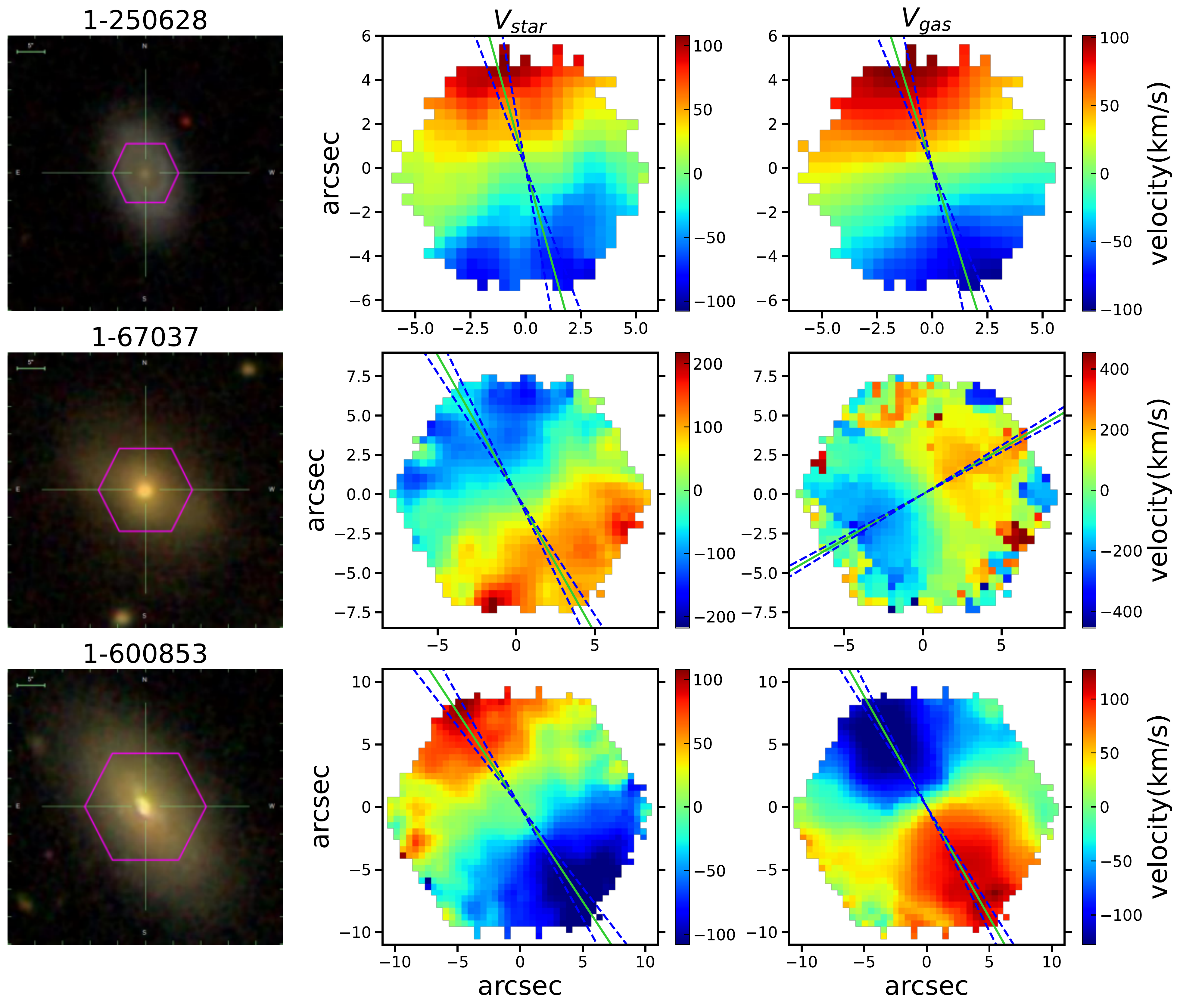}
    \caption{Examples of MaNGA galaxies with different types of kinematically misalignment. Each row corresponds to a MaNGA galaxy. The left column shows the SDSS $g$, $r$, $i$-band images with hexagons marking the regions of MaNGA bundle. The middle and right columns show stellar and gaseous velocity fields (traced by H$\alpha$) respectively. Red side moves away from us while blue side approaches us. The solid green lines mark the position angle (PA) of each component while the blue dashed lines mark the $\pm1\sigma$ error of PAs. Top panel shows a galaxy in which stellar and ionized gas components are rotation in the same way. The middle panel shows a galaxy with gas-star perpendicularly rotating while the bottom panel displays a gas-star counter-rotator.}
    \label{fig:misaligned_sample}
\end{figure*}

From the 9,456 unique galaxies in MPL-10, we first select 9,291 galaxies without faults and critical failure in DRP or DAP marked by DAPDONE and DAPQUAL keywords. To get the robust measurement of gas velocity field, we first discard 2,333 "lineless" galaxies. The "lineless" galaxies are defined as objects with H$\alpha$ S/N lower than 3 for more than 90\% spaxels within $\sim$1.5$R_e$. The remaining 6,958 sources are referred as emission-line galaxies. We then use PaFit package in Python \citep{krajnovi_kinemetry_2006} to fit the position angle (PA) for stellar and gaseous velocity fields of each galaxy. Spaxels within $1.5R_e$ with median spectral S/N>3 per spaxel are used for fitting, where bad spaxels are removed based on the MaNGA DAP flag.

With the measurements of PA of stellar component ($\rm{PA}_{\rm{star}}$) and gaseous component ($\rm{PA}_{\rm{gas}}$), the position angle offset $\Delta \text{PA}\equiv|\rm{PA}_{\rm{star}}-\rm{PA}_{\rm{gas}}|$ is calculated for each emission-line galaxy. We then select 723 candidates of misaligned galaxies defined as $\Delta\mathrm{PA}>30^{\circ}$ with robust PA measurements (${\mathrm{PA}}_{\mathrm{error}}\leq60^{\circ}$, where $\mathrm{PA}_{\mathrm{error}}$ is the $1\sigma$ error of PAs).

Finally, we visually check the 723 sources and select 456 misaligned galaxies. On-going mergers, galaxies with irregular morphologies, groups within a single MaNGA bundle are removed through this process since it is hard to assign a major axis. In summary we obtain 456 misaligned galaxies from 6,958 emission-line galaxies in MPL-10.

Fig.~\ref{fig:misaligned_sample} shows three examples of MaNGA galaxies. The left column gives the false color image. The middle and right column display the stellar and H$\alpha$ velocity field respectively. The solid green lines mark the position angle (PA) of each component while the blue dashed lines mark the $\pm1\sigma$ error of PAs. The top row shows clear kinematic alignment and the middle row displays perpendicular gas-star misalignment. The bottom row shows an example of counter-rotator.

We cross match the MaNGA galaxies with catalog from \cite{chang_stellar_2015} and obtain the global stellar mass ($M_*$), star formation rate (SFR) for 6,217 emission-line galaxies and 416 out of 456 misaligned galaxies. The $M_*$ and SFR are estimated through photometric spectrum energy distribution (SED) fitting based on data from SDSS and WISE photometry with Chabrier initial mass function \citep[IMF,][]{chabrier_galactic_2003}. Fig.~\ref{fig:sampgal_SFR_M_relation} shows the SFR vs. $M_*$ diagram with grey dots representing MaNGA galaxies and the red squares being the misaligned sample. The black contours depict galaxies from \cite{chang_stellar_2015}. It is clear that there are two density peaks of these contours. The two black dashed lines \citep{jin_sdss_2016} separate galaxies into three populations, where the upper line is the $-1\sigma$ scatter of the star-forming main sequence and the lower line is $\log\mathrm{sSFR}\sim-15$ \citep{chen_growth_2016}. The top density peak with high SFR is defined as the star forming main sequence (SF), while the bottom peak with little star formation is the red quiescent sequence (QS). The population in between these two extremes is defined as green valley (GV). These 416 misaligned galaxies are divided into 74 SF, 136 GV and 206 QS galaxies. And the fractions of misaligned galaxies in emission-line galaxies with different galaxy types are 2.4\% (SF), 7.3\% (GV) and 15.6\% (QS) respectively.

\subsection{Control Samples}
\label{subsection:data_analysis}
\begin{figure*}
    \includegraphics[width=2.0\columnwidth]{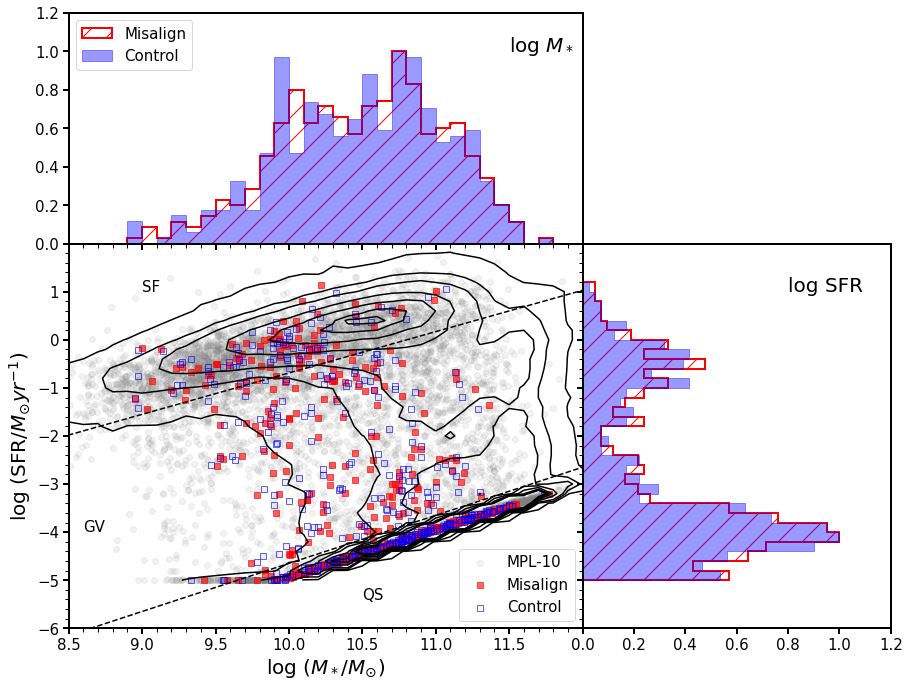}
    \caption{SFR versus $M_*$ diagram for kinematically misaligned galaxies and one of the control samples. The black contours represent galaxies from \protect\cite{chang_stellar_2015}. The grey open circles represent MaNGA MPL-10 galaxies with $M_*$ and SFR measurements in Chang's catalogue. Red squares represent misaligned galaxies while blue open squares represent control samples. Two dashed lines separate galaxies into star-forming, green-valley and quiescent. The top and right panels show distribution of stellar masses and SFR for the misaligned galaxies (red) and their control samples (blue) respectively. The peak of each distribution is set to 1.}
    \label{fig:sampgal_SFR_M_relation}
\end{figure*}
It is believed that the gas of the kinematically misaligned galaxies has an external origin like mergers and gas accretion. In order to quantify the influence of external gas acquisition on the following evolution of progenitor galaxies, we build ten non-misaligned control samples with $\Delta\rm{PA}<30^{\circ}$ for comparison. For each misaligned galaxy, we randomly select ten non-misaligned galaxies which are closely matched in the global $M_*$ and SFR, with $|\Delta(\log M_*)| < 0.1$ and $|\Delta(\log\rm{SFR})| < 0.2$. The motivation of choosing $M_*$ and SFR as matching parameters includes: i) stellar mass is the most fundamental property of a galaxy and tightly correlates with many other physical parameters; (ii) requiring the control galaxies to have similar SFR is extremely important since the fraction of kinematically misaligned galaxies strongly depends on star formation activity (see Section~\ref{subsection:incidence}). We show one of the control samples in Fig.~\ref{fig:sampgal_SFR_M_relation} as blue squares. The right-hand panel gives the distribution of SFR for misaligned (blue) galaxies and control (red) samples while the top panel gives the distribution of $M_*$. The peak of each distribution is set to 1. It is clear that the sample and controls have similar distributions in $M_*$ and SFR.

\section{Results}
\label{section:results}
\subsection{$\Delta$PA Distribution}
\begin{figure}
   \includegraphics[width=1.0\columnwidth]{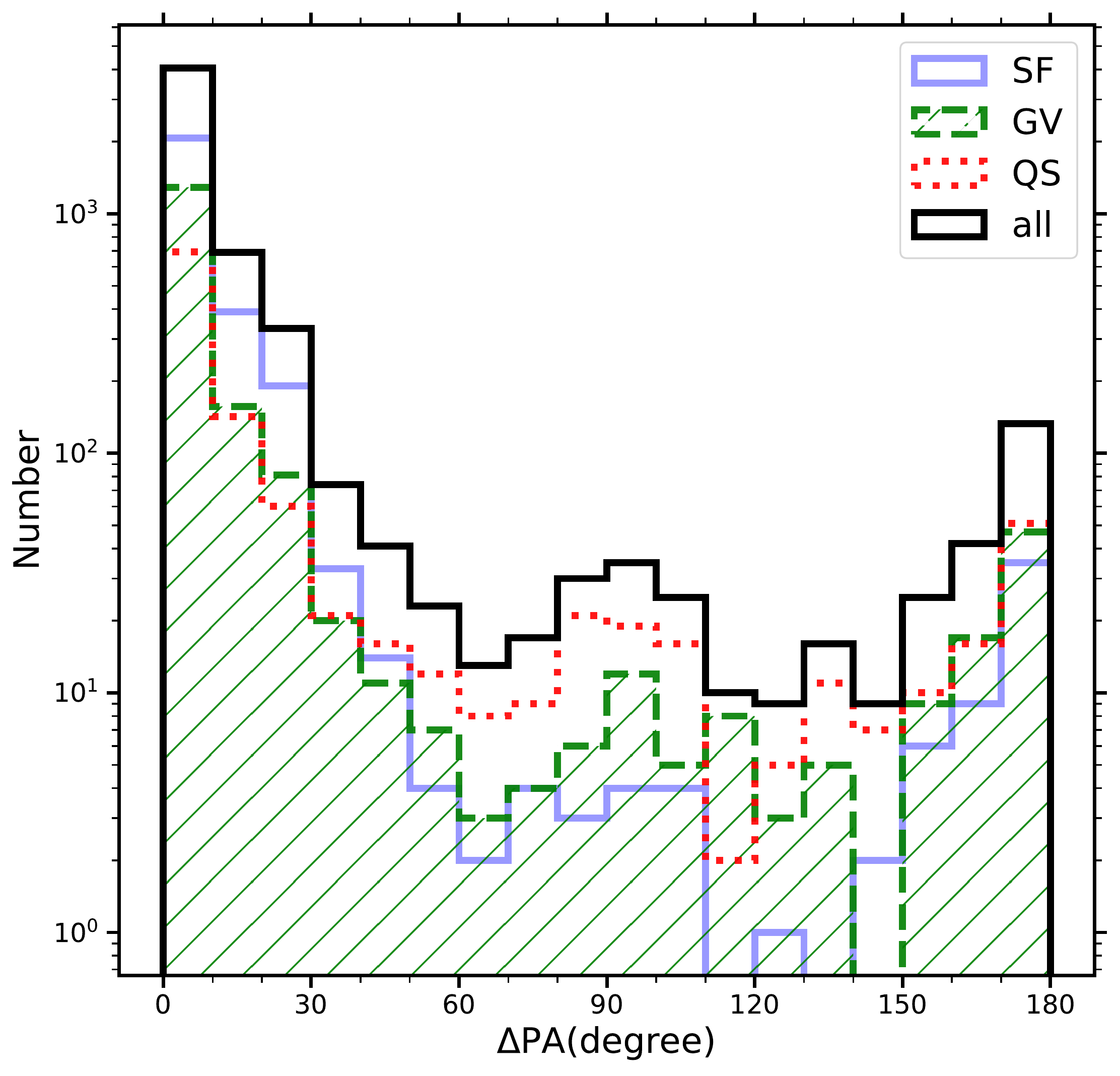}
    \caption{$\Delta$PA (position angle difference between stellar and gaseous velocity fields) distribution for all galaxies (black solid) and galaxies with different galaxy types as SF (blue solid), GV (green dashed) and QS (red dotted).}
    \label{fig:Delta_PA_distribution}
\end{figure}
Fig.~\ref{fig:Delta_PA_distribution} shows the $\Delta\mathrm{PA}$ distribution for galaxies with robust gaseous and stellar PA measurements (black solid). The blue solid, green dashed and red dotted histograms represent the SF, GV and QS samples respectively. There are three local peaks at $\Delta\mathrm{PA}\approx0^{\circ}$, $90^{\circ}$ and $180^{\circ}$. To quantify the statistical significance of the peak at $90^{\circ}$, we perform the one-sample KS test with respect to a normal distribution which has a mean value of $90^{\circ}$ with standard deviation of $15^{\circ}$. The $p$-value for SF, GV, QS and all galaxies are $0.347$, $0.192$, $0.552$ and $0.928$ respectively. The $p$-value larger than $0.05$ suggests that we cannot reject the hypothesis that they are drawn from the normal distribution at the peak of $90^{\circ}$. This triple peak distribution can be easily understood that the gas disks will feel a torque from stellar disks in case of $\Delta\mathrm{PA}\neq0^{\circ}$, $90^{\circ}$, $180^{\circ}$, which will pull the gas disk to co-rotating or counter-rotating with respect to stellar disks. 

\subsection{The fraction of misaligned galaxies}
\label{subsection:incidence}
\begin{figure*}
   \includegraphics[width=2.0\columnwidth]{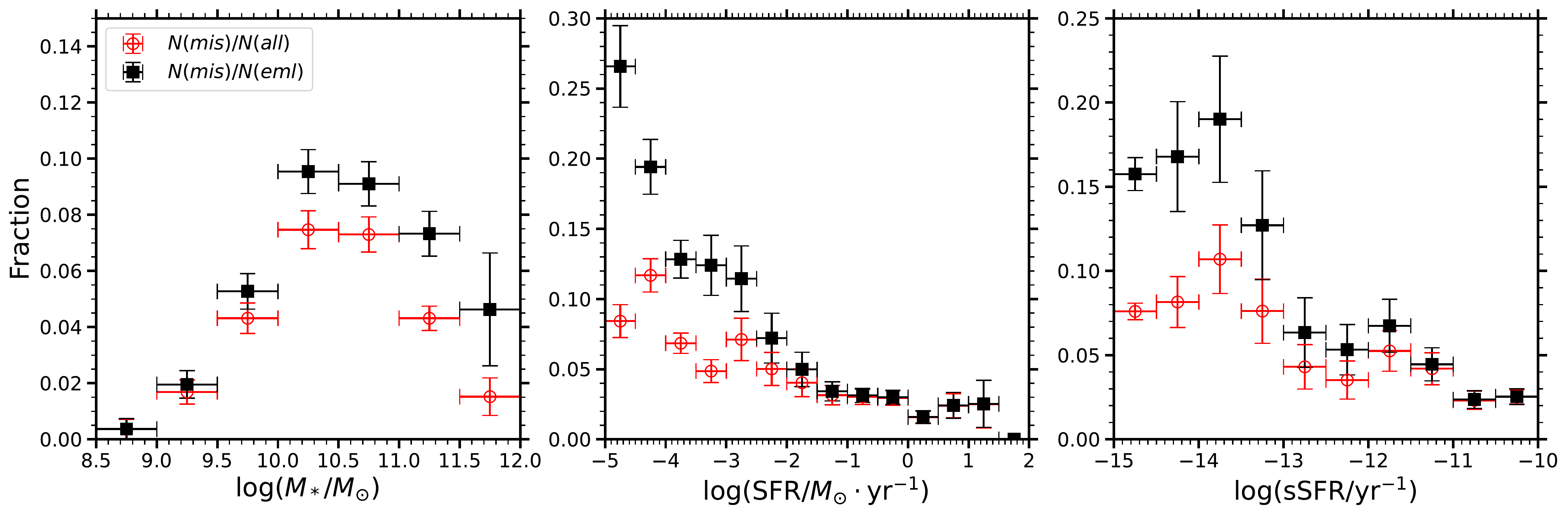}
    \caption{Fraction of misaligned galaxies as a function of $M_*$, SFR and sSFR. Black square represents misaligned fraction with respect to emission-line galaxies $\mathrm{N}(\mathrm{misalign})/\mathrm{N}(\mathrm{eml})$ and red circle represents misaligned fraction with respect to all galaxies $\mathrm{N}(\mathrm{misalign})/\mathrm{N}(\mathrm{all})$. The errorbar for $x$-direction shows the size of each parameter bin while the errorbar in $y$-direction is obtained by using bootstrap method.}
    \label{fig:misalign_fraction}
\end{figure*}

It is well known that the misaligned galaxies are ubiquitous in S0 and elliptical galaxies, but much fewer in blue star forming galaxies. This could be due to the large cross section between gaseous components. If misaligned galaxies accrete external gas, the newly acquired gas will collide with the pre-existing gas in the progenitors, leading to the redistribution of angular momentum \citep{chen_growth_2016}. Thus, in the gas-poor early type galaxies, the accreted misaligned gas would be easier to survive since the interaction with pre-existing gas is less important and the collisional cross-section between gas and star is too small to influence the angular momentum of the accreted gas, while in the gas-rich star forming galaxies, the misaligned phenomenon would only appear when the angular momentum of the accreted gas is larger than the pre-existing gas. 

In the MaNGA target selection strategy \citep{wake_sdss_2017}, there is no cuts on size, inclination, morphology or environment. It provides an unprecedented IFU sample which is fully representative of the local galaxy population and the selection is well understood. Based on the largest misaligned sample from MaNGA, we analyze the dependence of kinematic misalignment fraction on galaxy physical parameters ($M_*$, SFR, $\rm{sSFR}\equiv\rm{SFR}/\rm{M}_{*}$) over all the galaxy populations.

Fig.~\ref{fig:misalign_fraction} shows how the fraction of misaligned galaxies varies with  $M_*$ (left), SFR (middle) and sSFR (right). We estimate the errorbar by bootstrap method. The red circles are fractions of misaligned galaxies with respect to all galaxies (including emission-line and "lineless" galaxies) in a certain parameter bin, while black squares are fractions with respect to only emission-line galaxies. We find that the misaligned fraction peaks at $\log(M_*/M_{\odot})\sim10.5$ and declines to both low and high mass end. This fraction also decreases monotonically with increasing SFR and sSFR. In essence, this is the dependence on the amount of pre-existing gas. Galaxies with higher star formation activity are expected to contain larger amount of pre-existing gas, which makes it difficult for the newly accreted gas to maintain misaligned angular momentum. The decline of this fraction at the low mass end is primarily due to the fact that the low mass galaxies with $\log(M_*/M_{\odot})<10\sim10.5$ in the local universe are dominated by star-forming galaxies with higher sSFR. At the high mass end, the decrease of this fraction is consistent with \cite{davis_atlas_2011} and \cite{jin_sdss_2016}. The low fraction of misalignment in the massive galaxies is suggested to be caused by galactic scale processes which suppress the accretion of external cold gas onto the galaxy, such as AGN feedback, existence of a hot X-ray gas halo, or a halo mass threshold.

\subsection{Global properties of misaligned galaxies}
\subsubsection{Kinematic asymmetry}
\begin{figure*}
    \includegraphics[width=2.0\columnwidth]{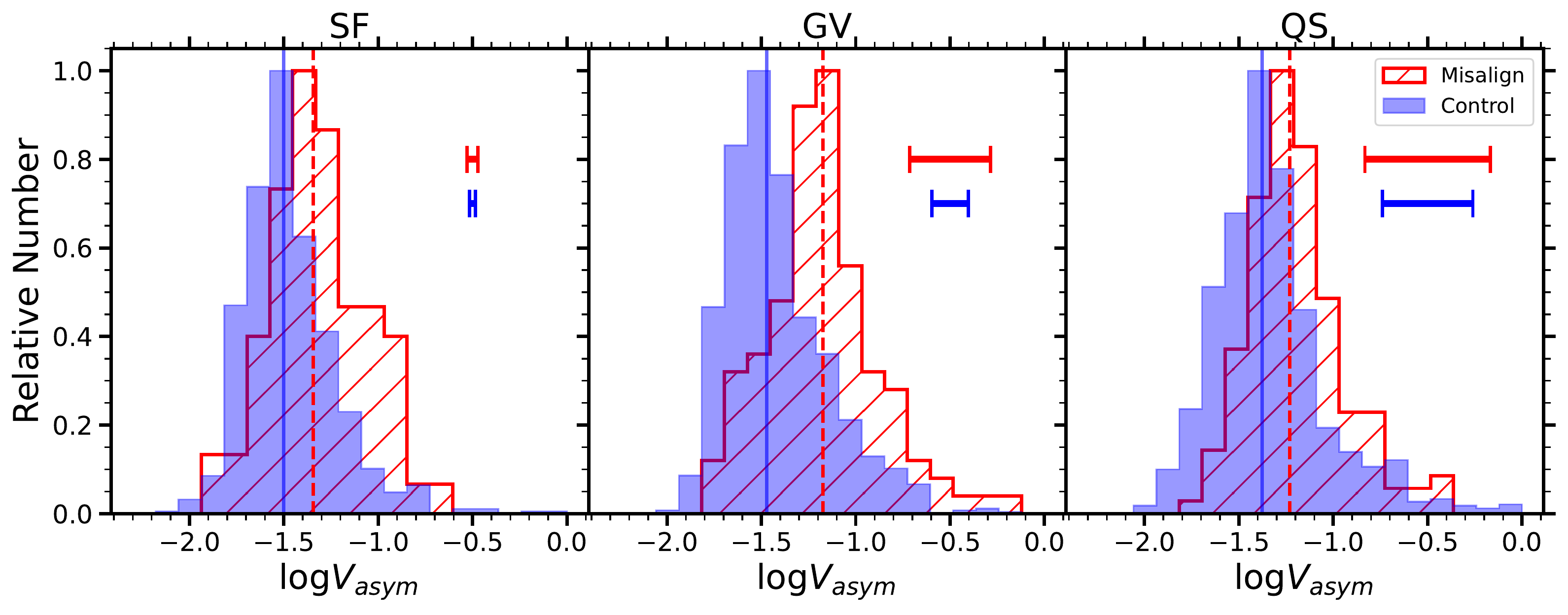}
    \caption{Kinematic asymmetry distribution of the misaligned galaxies (red) and their control samples (blue) for three galaxy types (SF, GV and QS). The peak of each distribution is set to 1. The dashed red and solid blue lines mark the median value of each distribution with the same color. The typical errors are shown in the top-right of each panel for misaligned galaxies (red) and their controls (blue).}
    \label{fig:VAsym_distribution}
\end{figure*}
Considering that the misaligned galaxies have gone through gas accretion, external interaction or merging process, we would expect to find evidence of disturbance of gas velocity fields. In this section, we compare the asymmetry parameters ($V_{\mathrm{asym}}$) of gas velocity fields between misaligned galaxies and their control samples.

$V_{\mathrm{asym}}$ used in this work is defined exactly the same as Equation (2) in \cite{feng_sdss_2020} and basically calculates higher-order coefficients of Fourier components normalized by the first-order coefficient, where the higher-order coefficients describe the asymmetric pattern of the velocity fields and the first-order represents symmetric patterns of the velocity fields.  Therefore higher value of $V_{\mathrm{asym}}$ means higher asymmetry. We refer readers to \cite{feng_sdss_2020} for more details of asymmetry parameter estimation.

Fig.~\ref{fig:VAsym_distribution} shows the distribution of kinematic asymmetry of gaseous component for the three galaxy types: SF (left), GV (middle) and QS (right). The red and blue histograms represent kinematically misaligned galaxies and their control samples, respectively. The peak of each distribution is set to 1. The dashed red and solid blue lines mark the median value of each distribution for misaligned galaxies and their controls, while the red and blue lines shown in top-right represent the median error of $\log V_{\mathrm{asym}}$ for misaligned galaxies and control samples respectively. The diagram shows that in SF and GV galaxies, the asymmetry of gas velocity fields is higher in misaligned galaxies than their control samples, which is totally consistent with our expectation that the acquisition of external gas would cause larger asymmetry in gaseous components of misaligned galaxies. For QS misaligned galaxies, we do not have definite conclusion due to the larger error, which can be attributed to the S/N of H$\alpha$ flux. 

\subsubsection{Gaseous content}
\label{subsection:HI}
It is suggested that misaligned galaxies acquire gas from external environments, which would lead to higher amount of gas in misaligned galaxies. On the other hand, the interaction between the pre-existing and accreted gas will cause redistribution of angular momentum and gas inflow, resulting in a rapid transformation of gas into stars. This process is especially important in the SF misaligned galaxies, in which their star-formation activity within central 1kpc is several times higher than the average of local star-forming galaxies \citep{chen_growth_2016}. From this point of view, the cold gas fraction could be lower in the misaligned galaxies since they are transformed into stars efficiently. We do not have cold gas observations for the sample. However, HI-MaNGA program provide GBT \ion{H}{i} observation for a subsample. In this section, we compare the {\ion{H}{i}} detection rate and molecular gas mass fraction estimated from dust extinction \citep{barrera_sdss_2018} between misaligned galaxies and their control samples.

HI-MaNGA is a follow-up program to detect {\ion{H}{i}} for MaNGA galaxies using the Green Bank Telescope (GBT). The third data release of HI-MaNGA value-added catalogue contains 6,632 galaxies, where 3,358 of them are observed by GBT and the remaining galaxies are within ALFALFA catalogue \citep{haynes_arecibo_2018}. The observation depth for GBT is about $1.5~\mathrm{mJy}$ and velocity resolution is $\sim10~\mathrm{km}\cdot\mathrm{s}^{-1}$. We refer readers to \cite{masters_HIMaNGA_2019} \& \cite{stark_HI_2021} for more details about the HI-MaNGA project.

\begin{figure}
    \includegraphics[width=1.0\columnwidth]{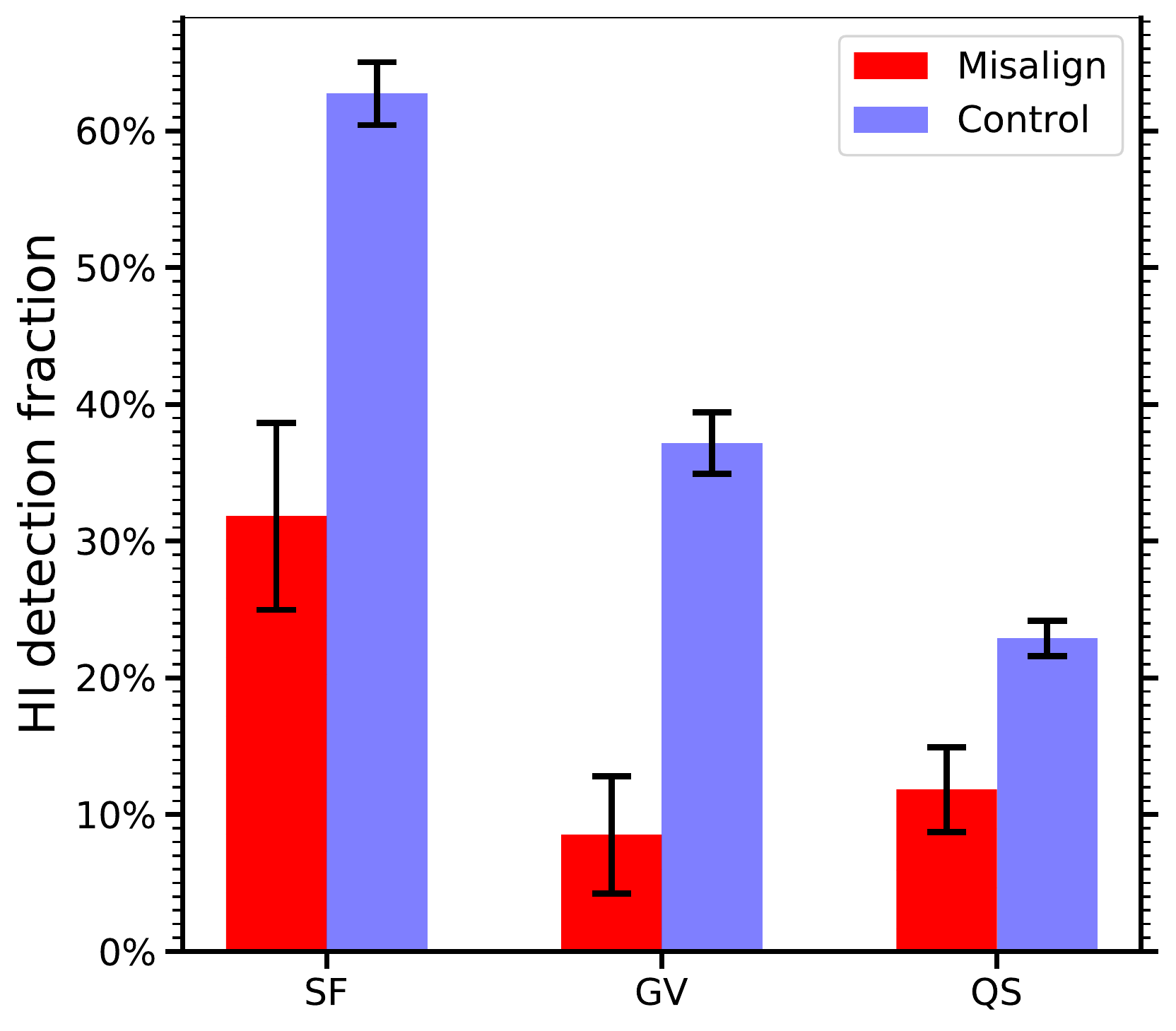}
    \caption{{\ion{H}{i}}-detection rate (${N_{\mathrm{detected}}}/{N_{\mathrm{observed}}}$) for misaligned galaxies (red) and their control samples (blue) classified into three galaxy types (SF, GV and QS).}
    \label{fig:HIdet_fraction}
\end{figure}

We cross-match the misaligned galaxies with HI-MaNGA value-added catalogue, finding 302 out of 416 matches and 90 of them are detected. Considering the beam size of GBT and ALFALFA is in the order of arcmin, the {\ion{H}{i}} detection could be contaminated by nearby galaxies. \cite{stark_HI_2021} flag potential confused sources which are located within 1.5 times the half-power beam width (HPBW) and at similar redshift of the primary target. They also estimate the probability $P_{R>0.2}$ where $R$ represents the fraction of contribution to the total measured flux from the companions. In this work, we remove 44 sources in which \ion{H}{i} flux is contaminated by limiting confusion flag $= 1$ and $P_{R>0.2}>0.1$ \citep{stark_HI_2021}. We build 10 control samples from \ion{H}{i}-observed galaxies for the 258 ($=302-44$) misaligned galaxies. The control samples are selected in a similar way as described in Section~\ref{subsection:data_analysis} but with an additional requirement: $\left|\Delta  z\right|<0.001$. Fig.~\ref{fig:HIdet_fraction} shows the {\ion{H}{i}}-detection rate (${N_{\mathrm{detected}}}/{N_{\mathrm{observed}}}$) for the misaligned galaxies and control samples, where $N_{\mathrm{observed}}$ is the number of galaxies matched in HI-MaNGA catalogue and $N_{\mathrm{detected}}$ is the number of galaxies detected in {\ion{H}{i}}. Red histograms in Fig.~\ref{fig:HIdet_fraction} are for the misaligned sample and blue for their controls. It is clear that the \ion{H}{i}-detection rate is significantly lower in misaligned galaxies than their controls for SF, GV and QS. 

\begin{figure*}
    \includegraphics[width=2.0\columnwidth]{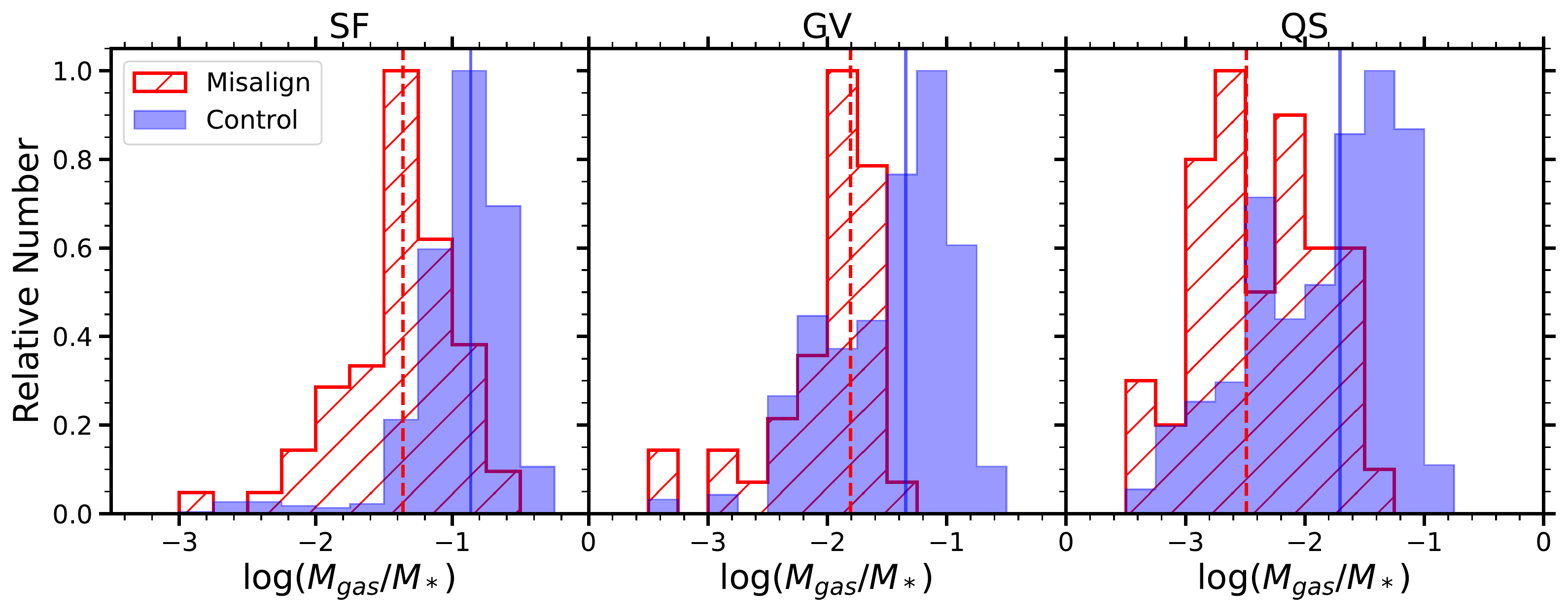}
    \caption{Molecular gas mass fraction ($M_{\mathrm{gas}}/M_*$) distribution of misaligned galaxies (red) and their control samples (blue) for three galaxy types (SF, GV and QS). The peak of each distribution is set to 1. The dashed red and solid blue lines mark the median value of each distribution with the same color.}
    \label{fig:Ha_mass_frac_distribution}
\end{figure*}

Considering that there are only 46 ($=90-44$) \ion{H}{i}-detected misaligned galaxies, this sample is hardly to give a robust statistics on \ion{H}{i} mass distribution. We estimate the molecular gas mass $M_{\mathrm{gas}}$ based on the method described in \cite{barrera_sdss_2018}. The molecular gas mass is estimated as $\Sigma_{\mathrm{gas}}=30M_{\odot}\mathrm{pc}^{-2}A_V$, where optical extinction $A_V$ is from the Balmer decrement. We construct 10 control samples with the same criteria as we construct control samples for \ion{H}{i}-observed misaligned galaxies. Fig.~\ref{fig:Ha_mass_frac_distribution} shows the distribution of molecular gas mass fraction defined as $M_{\mathrm{gas}}/M_*$  between misaligned galaxies (red) and control samples (blue). It is clear that misaligned galaxies have lower molecular gas mass fraction than the control samples.

A similar result is reported in \cite{duckworth_decouplingI_2020} based on a sample of $\sim$4500 galaxies in MaNGA MPL-8, where the gas mass is estimated in exactly the same way. Based on IllustrisTNG simulation, \cite{lu_hot_2021} show that low mass counter-rotating disc galaxies possess smaller {\ion{H}{i}} mass fraction within twice the half stellar mass radius $2R_{\rm{hsm}}$. The evidence of lower {\ion{H}{i}} detection rate and molecular gas mass fraction in the SF and GV misaligned sample could be the consequence of high star-formation effieciency ($\mathrm{SFE}=\mathrm{SFR}/M_{\mathrm{gas}}$). Based on the MaNGA survey, \cite{chen_growth_2016} compare the star-formation activity parameter $\alpha_{\mathrm{SF}}=1/[\mathrm{sSFR}\times(t_\mathrm{H}(z)-1\mathrm{Gyr})]$ between SF counter-rotators and the average of local SF galaxies, where $t_{\mathrm{H}}(z)$ is the Hubble time at redshift $z$ and $\alpha_{\mathrm{SF}}$ compares the current SFR and the past average. They show that the central star-formation activity of SF misaligned galaxies is several times higher than the average of local SF galaxies within central 1kpc. Based on a larger sample, \cite{xu_sdss_2022} discover that the central region of misaligned SF galaxies have higher sSFR in comparison with their control samples, and the misaligned galaxies shows smaller central $\mathrm{D}_n4000$ indicating younger stellar populations in SF and GV misaligned galaxies. For the QS misaligned galaxies, the {\ion{H}{i}} and molecular gas deficiency could be due to a selection bias that progenitors with lower gaseous contents are easier to show kinematic misalignments, since the lack of interaction between accreted and pre-existing gas.

\subsubsection{Size}
\label{subsection:size}
\begin{figure*}
    \includegraphics[width=2.0\columnwidth]{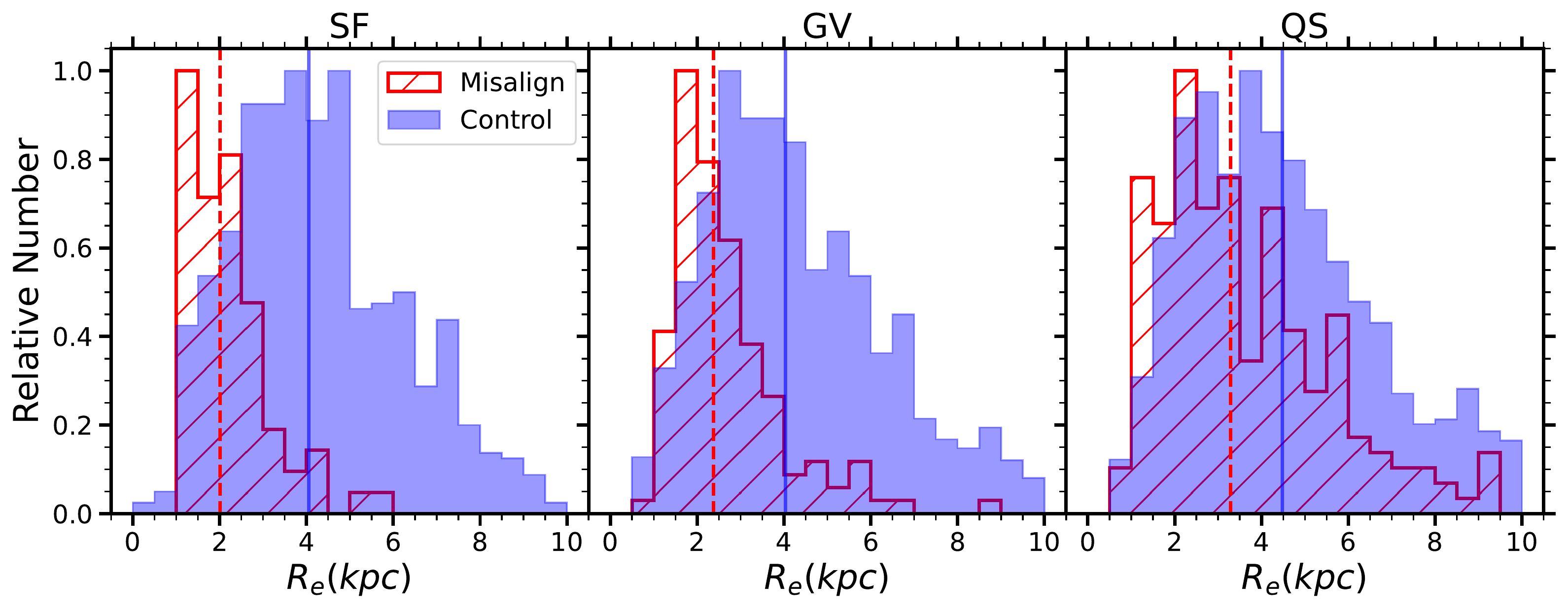}
    \caption{Effective radius distribution of the misaligned galaxies (red) and their control samples (blue) for three galaxy types (SF, GV and QS). The peak of each distribution is set to 1. The dashed red and solid blue lines mark the median value of each distribution with the same color.}
    \label{fig:Re_distribution}
\end{figure*}
As suggested by \cite{chen_growth_2016}, gas-gas collision leads to fast gas inflow, triggering the central star formation. These newly formed stars may lead to the change of stellar light/mass distribution, especially for the gas-rich SF galaxies. In this section, we compare $r$-band effective radius $R_e$, which is defined as a radius containing 50\% emission of a galaxy, between misaligned galaxies and their control samples.

In this work, we take $R_e$ from NASA-Sloan Atlas (NSA)\footnote{\url{http://nsatlas.org/}} \citep{blanton_improved_2011} and convert it to physical scale in unit of kpc. Fig.~\ref{fig:Re_distribution} shows the distribution of $R_e$ for the misaligned galaxies (red) and control samples (blue). The red and blue vertical lines mark the median value of each distribution. There is a clear trend that misaligned galaxies have smaller $R_e$ than the control samples for all the three types of galaxies. It is about $1.5-2~\mathrm{kpc}$ smaller for SF and GV misaligned galaxies than their controls and $1~\mathrm{kpc}$ smaller for QS misaligned galaxies.

This result is within our expectation under the picture of angular momentum loss due to gas-gas collision for the SF galaxies, and then the in-falling gas gathers in the central region triggering star formation, which increases the brightness at the galaxy center. Since $R_e$ quantifies the radius at which half of light is emitted, the central star formation would be manifested as the decline of $R_e$. Similar results are also found in simulation. \cite{starkenburg_origin_2019} find that counter-rotating galaxies with $\Delta\mathrm{PA}>90^{\circ}$  have smaller half stellar mass radius compared to the whole sample with $9.3<\log(M_*/M_{\odot})<10.7$. \cite{lu_hot_2021} discover the star-forming counter-rotators with $9.7<\log(M_*/M_{\odot})<10.3$ also have smaller half stellar mass radius compared to normal disc galaxies. The difference in $R_e$ between GV misaligned galaxies and their controls indicates that similar process happen in GV galaxies. The difference in $R_e$ between QS misaligned galaxies and their controls is smaller, we will discuss this difference in Section \ref{subsection:discussion of QS}.

\subsubsection{Morphology}
\begin{figure*}
    \includegraphics[width=2.0\columnwidth]{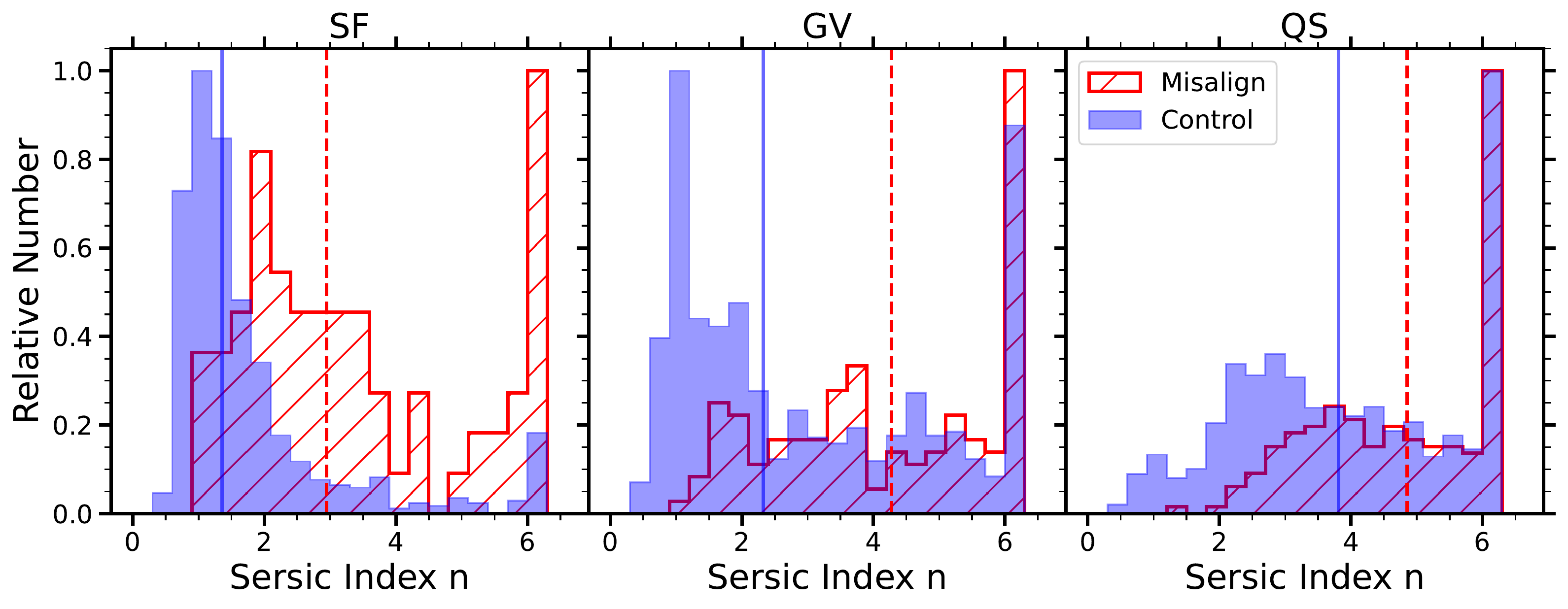}
    \caption{Sersic index $n$ distribution of misaligned galaxies (red) and their control samples (blue) for three galaxy types (SF, GV and QS). The peak of each distribution is set to 1. The dashed red and solid blue lines mark the median value of each distribution with the same color.}
    \label{fig:Sersic_index_distribution}
\end{figure*}
Galaxy morphology can be characterized by the S\'{e}rsic index. The surface brightness profile can be described as:
\begin{equation}
I(R)=I_0\cdot\exp\left[-\beta_n\cdot\left(\frac{R}{R_e}\right)^{1/n}\right],
\end{equation}
where $I(R)$ is surface brightness at radius $R$, $I_0$ is central surface brightness and $n$ is called S\'{e}rsic index. The case of $n=2$ is used as the demarcation for disc dominated versus bulge dominated galaxies. The value of S\'{e}rsic index for MaNGA galaxies is taken from the NASA-Sloan Atlas catalogue. Fig.~\ref{fig:Sersic_index_distribution} shows the distribution of S\'{e}rsic index for the misaligned galaxies (red) and their control samples (blue). The red and blue vertical lines mark the median value of each distribution. 

For all the three types of galaxies, the misaligned galaxies show larger S\'{e}sic index than their control samples, with smallest difference in $n$ for QS galaxies. The median value for SF misaligned galaxies is $n\sim3$, while the median value for control sample is less than 2, implying SF misaligned galaxies are more bulge dominated compared to the disc nature of their controls. The GV misaligned galaxies (median $\sim4$) also show a more bulge dominated feature than controls (close to disc dominated, with median $\sim2$). Both QS misaligned galaxies and their controls display bulge dominated feature with median of $n$ for misaligned galaxies is a little bit higher than controls. As suggested by \cite{chen_growth_2016}, for SF misaligned galaxies, the redistribution of angular momentum happens during the interaction between accreted and pre-existing gas, leading to gas inflow and central star formation. The central star-formation will alter the light distribution, leading to higher S\'{e}rsic index of SF misaligned galaxies. The higher S\'{e}rsic index of GV galaxies could be attributed to a similar process.

\subsubsection{Spin parameter $\lambda_{R_e}$}
Based on the picture described in the previous sections, the interaction between accreted and pre-existing gas will lead to the loss of angular momentum. The newly formed stellar components from the gas will have lower angular momentum. To explore the kinematic state of stellar component of a galaxy, we calculate the spin parameter $\lambda_{R_e}$ as a proxy for the specific angular momentum of stars in the galaxy \citep{emsellem_sauronIX_2007}. This parameter is similar to $V/\sigma$ but incorporates spatial information, which is defined as
\begin{equation}
\lambda_R=\frac{\sum^{N_p}_{i=1}F_iR_i|V_i|}{\sum^{N_p}_{i=1}F_iR_i\sqrt{V_i^2+\sigma_i^2}}
\end{equation}
where $R$ is the radius within which the summation has been taken place. $F_i$, $v_i$, $\sigma_i$ are the $r$-band flux, stellar velocity and stellar velocity dispersion of the $i$th spaxel. $R_i$ is the distance between the $i$th spaxel and the galaxy center.

\begin{figure*}
    \includegraphics[width=2.0\columnwidth]{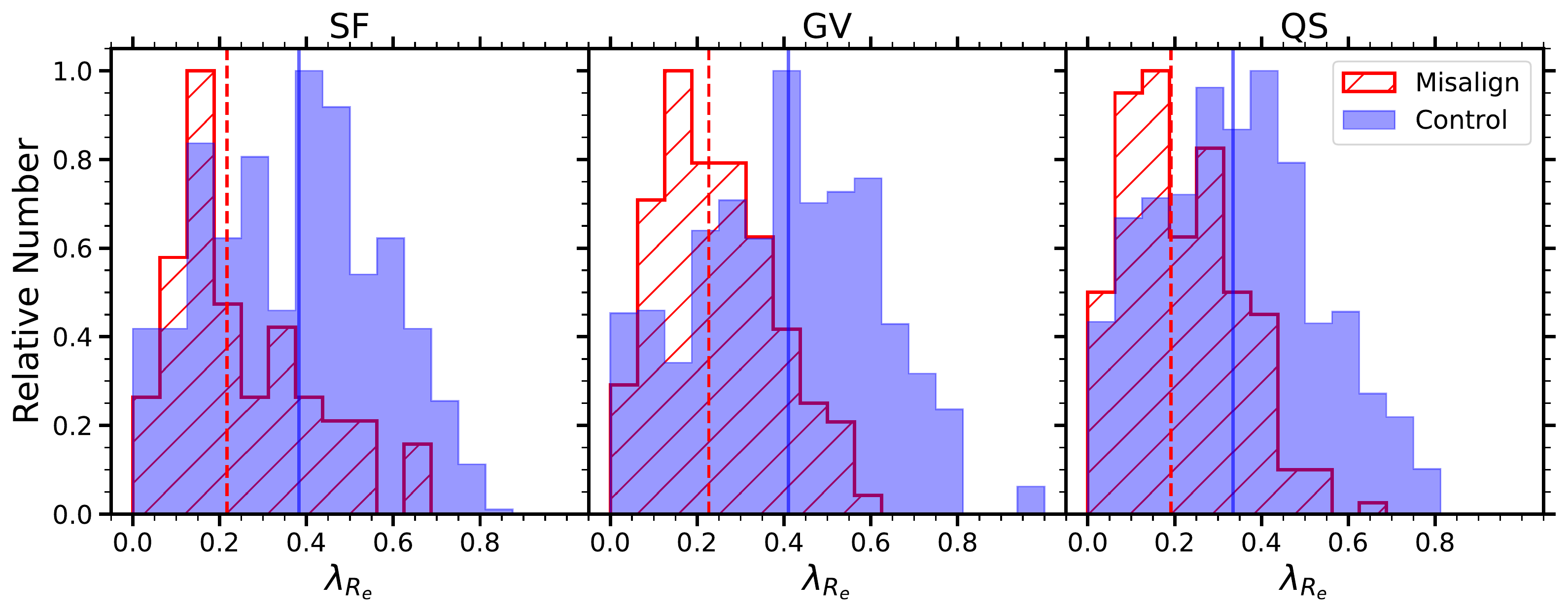}
    \caption{$\lambda_{R_e}$ distribution of misaligned galaxies (red) and control samples (blue) for three galaxy types (SF, GV and QS). The peak of distribution are set to 1. The dashed red and solid blue lines mark the median value of each distribution in the same color.}
    \label{fig:lambda_Re_distribution}
\end{figure*}

Fig.~\ref{fig:lambda_Re_distribution} show the distribution of $\lambda_{R_e}$ for the stellar components between misaligned galaxies (red) and control samples (blue). The two vertical lines mark the median value of each distribution. As $\lambda_{R_e}$ is a projected quantity, we add an additional control parameter inclination $\varphi_{\mathrm{incl}}$ with $|\Delta\varphi_{\mathrm{incl}}|<5^{\circ}$. Clearly, all the three types of misaligned galaxies tend to rotate slower than the control samples. Similar results are reported in both observations and simulations. \cite{duckworth_decouplingI_2020} find that misaligned galaxies with different morphological types in MaNGA survey are all have lower angular momentum in stellar components compared to their aligned counterparts, consistent with their results from IllustrisTNG simulation. \cite{starkenburg_origin_2019} show that gas-star counter-rotating galaxies display smaller stellar spin compared to a whole sample with $9.3<\log(M_*/M_{\odot})<10.7$. For SF misaligned galaxies, the lower $\lambda_{R_e}$ in misaligned galaxies is due to the loss of angular momentum during the interaction between accreted and pre-existing gas, leading to the newly formed stellar components with lower angular momentum. For GV misaligned galaxies, similar formation can be applied.

\subsubsection{Environment}
In the gas accretion scenario, we would expect the misaligned galaxies reside in environments which could provide the convenience for cold gas accretion. In dense environment, the gas stripping or halo heating mechanism may inhibit such process. From $\mathrm{ATLAS}^{\mathrm{3D}}$ survey, \cite{davis_atlas_2011} show early-type galaxies in dense group and cluster are more likely displaying kinematic alignments between gas and stars, while misalignments tend to exist in field galaxies. Based on 66 misaligned galaxies with different star-formation types in MaNGA survey, \cite{jin_sdss_2016} show that they all reside in more isolated environment. With a $\sim$10 times larger sample size, we revisit the environmental dependence of different types of misaligned galaxies in this section. We compare the environmental parameters between misaligned galaxies and their control samples.

We use two parameters to characterize the environment of misaligned galaxies, the neighbour count (Nneigh) and the tidal strength parameter ($Q_{\rm{lss}}$).
Nneigh is defined as the number of neighbours with $r$-band absolute magnitude brighter than $-19.5$ within $500~\mathrm{km}~\mathrm{s}^{-1}$ line-of-sight velocity and $5~\mathrm{Mpc}$ projected distance in a volume limited sample with $z<0.15$. The tidal strength parameter $Q_{\mathrm{lss}}$ is a description of the total gravitational interaction strength by all the neighbours within a fixed volume. The estimation method is described in \cite{argudo_catalogues_2015} and the result is released in the GEMA-VAC catalogue.
It is defined as:
\begin{equation}
Q_{\rm{lss}}=\log\left[\sum_i\frac{M_i}{M_p}\left(\frac{D_p}{d_i}\right)^3\right]
\end{equation}
where $M_i$ and $M_p$ are the stellar mass of the $i$th galaxy and the primary galaxy. $d_i$ is the projected distance between the primary galaxy and the $i$th neighbour with $D_p$ being the diameter of the primary galaxy.
\begin{figure*}
    \includegraphics[width=2.0\columnwidth]{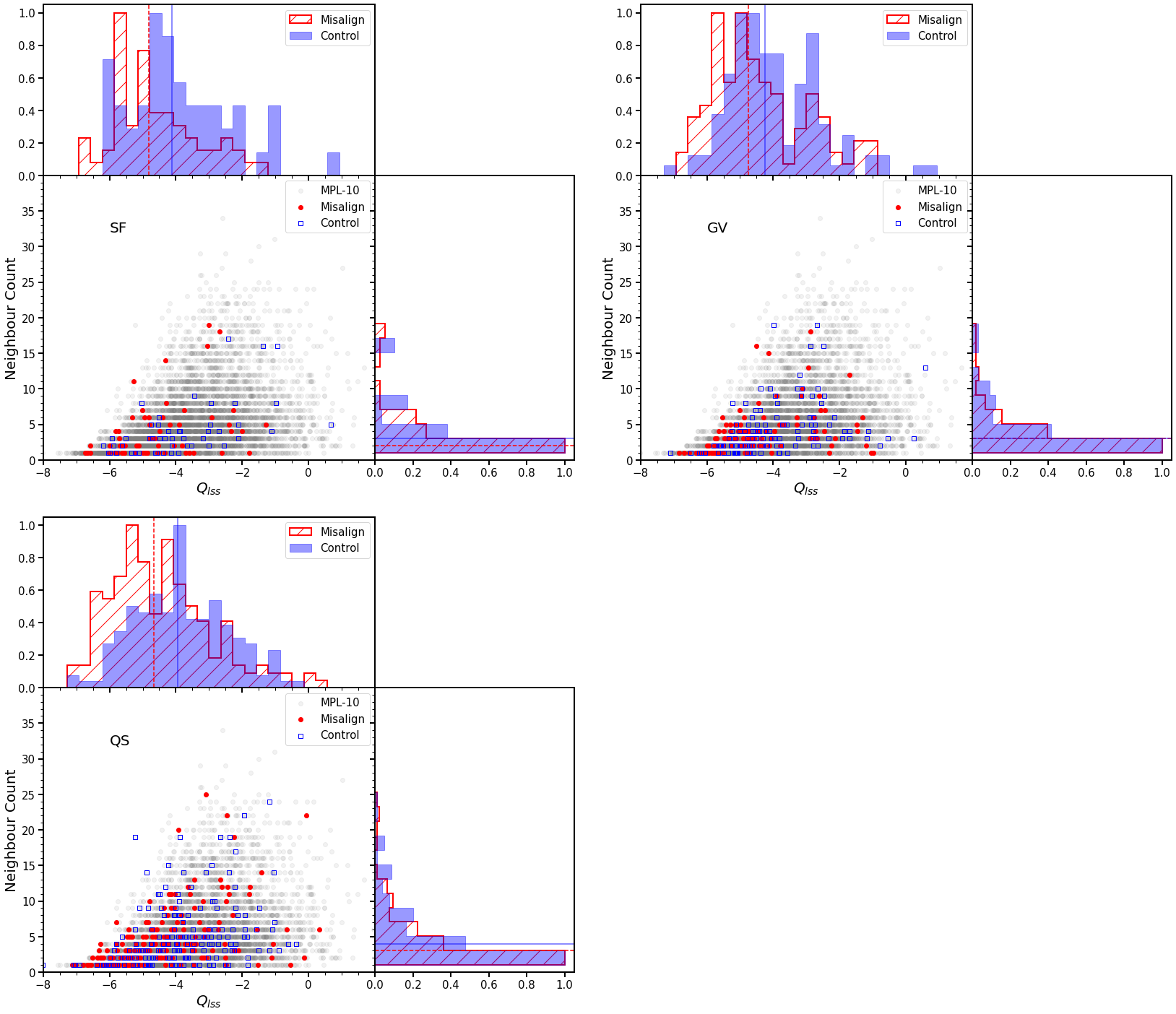}
    \caption{The neighbour count (Nneigh) versus tidal strength parameter $Q_{\mathrm{lss}}$ diagram for SF (top-left), GV (top-right) and QS (bottom-left) misaligned galaxies (red circle) and their control samples (blue square). The grey circles mark the MPL-10 galaxies. The top left and bottom right inset of each panel are the distribution of $Q_{\mathrm{lss}}$ and Nneigh for the misaligned galaxies and their controls respectively. The two vertical lines mark the median value of each distribution with the same color.}
    \label{fig:Qlss_nneigh}
\end{figure*}

Fig.~\ref{fig:Qlss_nneigh} shows Nneigh versus $Q_{\mathrm{lss}}$ diagram for SF (top-left), GV (top-right) and QS (bottom-left) misaligned galaxies (red) and their control samples (blue). The top left and bottom right insets of each panel are the distribution of $Q_{\mathrm{lss}}$ and Nneigh respectively. The two vertical lines mark the median value of each distribution.

As we can see from all the three types of galaxies, misaligned galaxies tend to reside in more isolated environment with lower value of $Q_{\mathrm{lss}}$ than their control samples. Similar trend is also found in Nneigh, but with much less significance.

\section{Discussion}

\subsection{Formation mechanisms of misaligned galaxies}
\label{subsection:discussion of QS}
It is believed that misaligned galaxies mainly form through external process, but the origin of external gas is still under debate. Galaxy interaction \citep{casanueva_origin_2021}, major/minor merger \citep{khim_star_2021} and gas accretion \citep{taylor_origin_2018, starkenburg_origin_2019, duckworth_decouplingI_2020,  duckworth_decouplingII_2020, khoperskov_extreme_2021} are all suggested to be the driver of kinematic misalignments.

In this work, we suggest different types of misaligned galaxies have undergone different formation mechanisms. For SF and GV misaligned galaxies, we propose a scenario that the progenitor galaxy accretes gas from gas-rich dwarf or cosmic web, followed by angular momentum redistribution through gas-gas interactions, leading to the star-formation in the central region. This toy model can naturally explain all of our observation results in section~\ref{section:results}. On the one hand, a large amount of SF and GV misaligned galaxies show S0-like morphologies. Numerical simulation from \cite{barnes_transformations_1992} suggest that merger, especially major merger could hardly keep the disk structure in spiral or S0 galaxies thus merger could not be a primary formation pathway for SF and GV misaligned galaxies. \cite{casanueva_origin_2021} study galaxies with $\log(M_*/M_{\odot})>9$ in EAGLE simulation, finding that contribution from mergers, including minor mergers, for the formation of misaligned galaxies is less than 5\%. Therefore, gas accretion is a more preferred mechanism in the formation of kinematic misalignments in SF and GV galaxies. On the other hand, both SF and GV misaligned galaxies show lower $D_n4000$ (younger stellar population) and higher specific SFR in the central regions \citep{chen_growth_2016, xu_sdss_2022}, indicating enhanced central star-formation activities.

We propose three possibilities to explain the origin of QS misaligned galaxies: (i) the QS misaligned galaxy is an evolutionary result of the GV misaligned galaxy. But this should not be the dominated formation mechanism because the number of GV misaligned galaxies is much less than their QS counterparts; (ii) mergers. \cite{li_impact_2021} find that the fraction of galaxies with the merging remnant features is $\sim$18\% in quiescent misaligned galaxies and $\sim$8\% in their co-rotating counterparts, while the remnant fractions for misaligned star-forming galaxies and their co-rotating counterpart are almost identical. \cite{khim_star_2021} study galaxies with $\log(M_*/M_{\odot})>10$ in Horizon-AGN simulation and point out the 35\% misaligned galaxies form through major and minor mergers; (iii) misaligned gas accretion. The observed differences in Section~\ref{section:results} between misaligned QS galaxies and their controls could be due to selection bias. An old progenitor galaxy tends to have higher possibility to be more bulge dominated and contain less cold gas fraction. It is easier for them to be observed as misaligned galaxies because the interaction between accreted and pre-existing gas can be neglected.

If we assume every galaxy has the same chance to accrete external gas, a misaligned fraction of $456/6958\approx7\%$ suggests that the kinematic misaligned phenomenon can last for $\sim$Gyr timescale ($1~\mathrm{Gyr}/13.7~\mathrm{Gyr}\approx7\%$). Based on the FIRE simulation, \cite{vandevoort_creation_2015} show that successive gas accretion can make the phenomenon of gas-star kinematic misalignments to last $\sim$Gyr, which is totally consistent with our observations.

\subsection{Lower gaseous content in misaligned galaxies}
The misaligned galaxies show lower {\ion{H}{i}} detection rate and molecular gas mass fraction. Similar results have been found for galaxies with different morphological types (ETGs, S0-Sas, Sb-Sd) in MaNGA survey \citep{duckworth_decouplingI_2020}, also in simulations for galaxies with $9.3<\log(M_*/M_{\odot})<10.8$ \citep{starkenburg_origin_2019}, dynamically hot disc galaxies with $9.7<\log(M_*/M_{\odot})<10.3$ \citep{lu_hot_2021}. 

One explanation for the lower gaseous content in SF and GV misaligned galaxies is that the cold gas is transformed into stars through higher star formation efficiency ($\mathrm{SFE}\equiv\mathrm{SFR}/M_{\mathrm{gas}}$). Using star-formation activity parameter $\alpha_{\mathrm{SF}}$, \cite{chen_growth_2016} show SF counter-rotators have several times higher star-formation activity in the central region compared to the average of local SF galaxies. Based on a larger sample, \cite{xu_sdss_2022} show that the SF misaligned galaxies have higher sSFR and lower $D_n4000$ (younger stellar population) in the central region than the control samples. Both \cite{chen_growth_2016} and \cite{xu_sdss_2022} describe a picture that the interaction between the pre-existing gas and the accreted gas leads to the angular momentum loss of gas. Thus the gas falls to the central region and triggers the star-burst/star-formation. The observed global properties in Section~\ref{section:results}, such as higher S\'{e}rsic index, smaller $R_e$, lower spin parameter $\lambda_{R_e}$, can naturally fit into this picture for the SF and GV misaligned galaxies.

Based on the Illustris simulation, \cite{starkenburg_origin_2019} and \cite{duckworth_decouplingI_2020, duckworth_decouplingII_2020} propose the low gaseous content for misaligned galaxies is due to gas loss from AGN feedback and flyby through group or cluster. \cite{starkenburg_origin_2019} discover $\sim$73\% galaxies show BH activity or flyby through a massive system like group or cluster within 1~Gyr before emergence of counter-rotation ($\Delta\mathrm{PA}>90^{\circ}$). \cite{duckworth_decouplingI_2020} find that no gas rotation galaxies (NGR galaxies, which cannot fit position angle for H$\alpha$ velocity fields due to gas depletion or dispersion dominated feature) and misaligned galaxies have similar stellar spin parameter ($\lambda_{1.5Re}$), which is lower than aligned galaxies, suggesting the progenitor galaxies experience gas removal (stage of no gas) and then re-accretion (stage of misalignment). \cite{duckworth_decouplingII_2020} show misaligned galaxies with $\log(M_*/M_{\odot})<10.2$ have boosted AGN luminosity, which leads to outflows and gas removal. 

However, the gas removal scenario faces difficulties in explaining the enhanced central star formation and the global properties like smaller $R_e$, lower spin parameter, higher S\'{e}rsic index observed in this work. On the other hand, \cite{khoperskov_extreme_2021} study 25 star-star counter-rotating galaxies with $\log(M_*/M_{\odot})<10.5$ in IllustrisTNG, finding that they all display gas-star misalignments due to gas accretion but without sudden gas loss by AGN feedback. We prefer the scenario that interaction between pre-existing and accreted gas triggers gas inflow, leading to enhanced central star formation and also fuelling the black hole, therefore displaying higher incidence of AGNs and boosted BH luminosity in misaligned galaxies than their controls. Both the activities of central BH and star-formation consume the cold gas, leading to lower cold gas content compared to their aligned counterparts.

For QS misaligned galaxies, the lower gaseous contents can be due to a selection bias, since progenitors with lower gas fraction would be easier to show kinematic misalignments.

\subsection{Beam smearing and inclination correction of $\lambda_{R_e}$}
Fig.~\ref{fig:lambda_Re_distribution} shows a large amount of SF misaligned galaxies with $\lambda_{R_e}<0.1$, indicating that they could be slow rotators. This result is totally out of our expectation. On the one hand, galaxies with higher S\'{e}rsic index and smaller $R_e$ suffer larger beam smearing effect \citep{harborne_recovering_2020}, leading to a lower $\lambda_{R_e}$ estimation. On the other hand, as $\lambda_{R_e}$ is a projected identity, the inclination will affect our measurement to a lower value. In this section, we check how the beam smearing and inclination effect influence our results on $\lambda_{R_e}$.

We correct the beam smearing effect based on Eqn.~(16) and Eqn.~(18) in \cite{harborne_recovering_2020}. They provide the corrections of $\lambda_{R_e}$ for galaxies with different $\sigma_{\mathrm{PSF}}/R_e$ and S\'{e}rsic index using a series of mock observations, where $\sigma_{\mathrm{PSF}}$ is the point spread function (PSF) of the observation. For SF galaxies, we adopt the inclination correction method described in \cite{fraser_sami_2021}. The inclination correction assumes that the intrinsic axial ratio is $q_0=0.2$ and the anisotropy parameter is $\beta=0.3$ for an edge-on disc galaxy, since SF galaxies are dominated by disc galaxies. The deprojected $\lambda_{R_e}$ satisfies
\begin{equation}
\lambda_{R_e}^{\mathrm{deproj}}=\frac{\lambda_{R_e}}{\sqrt{C^2-\lambda_{R_e}^2(C^2-1)}},
\end{equation}
where $C=\frac{\sin i}{\sqrt{1-\beta\cos^2i}}$, $\cos i=\sqrt{\frac{(b/a)^2-q_0^2}{1-q_0^2}}$ and $b/a$ is the observed axial ratio.

\begin{figure*}
    \includegraphics[width=2.0\columnwidth]{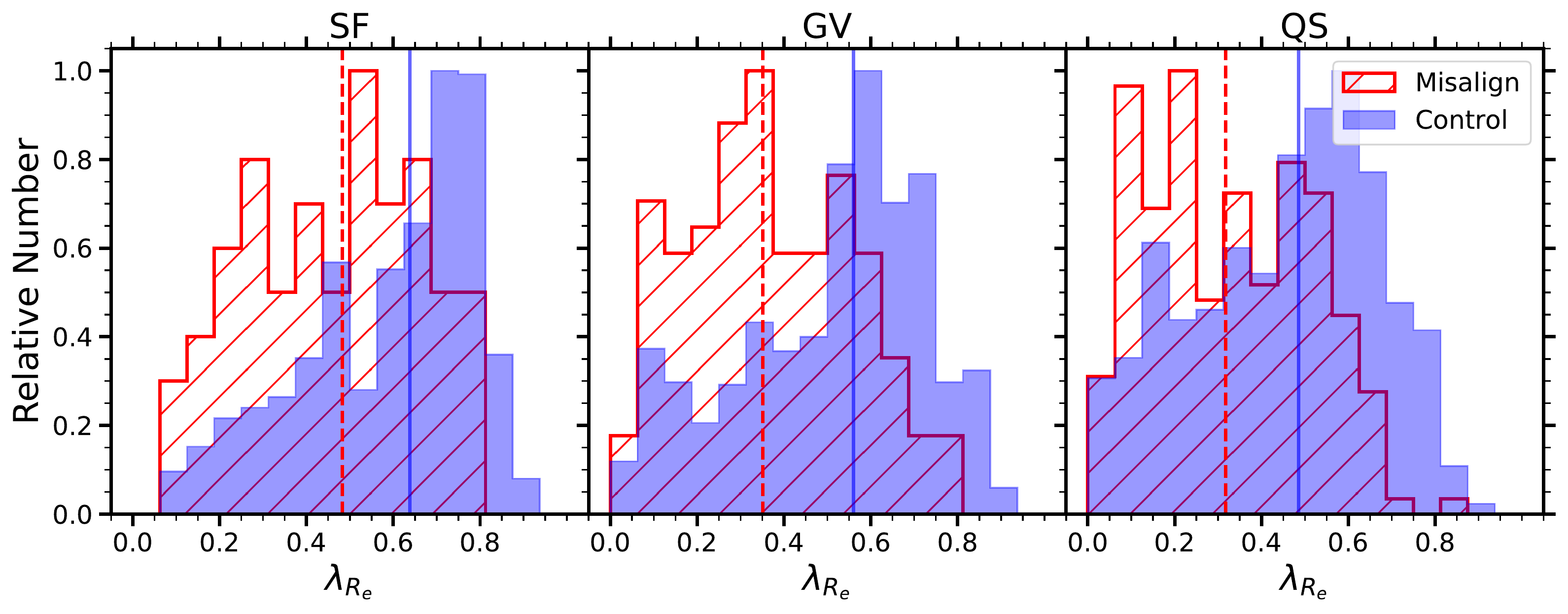}
    \caption{The $\lambda_{R_e}$ distribution betwee misaligned galaxies and control sample in three galaxy types (SF, GV and QS) after beam smearing and inclination correction.}
    \label{fig:comp_lambdaRe_corr}
\end{figure*}

Before any correction is applied, Fig.~3 of \cite{fraser_sami_2021} shows the $\lambda_{R_e}$ of SF galaxies ranges from 0.3 to 0.5 depending on their locations on the SF main-sequence. A median value of $\lambda_{R_e}\sim0.4$ for our SF control samples is consistent with Fraser-McKelvie's result. Once the beam smearing and inclination corrections are applied, the median value of $\lambda_{R_e}$ in SF control sample increases to $\sim$0.6 as shown in Fig.~\ref{fig:comp_lambdaRe_corr}, which is totally consistent with Fraser-McKelvie's result. The $\lambda_{R_e}$ of misaligned galaxies are smaller than their control samples in all the three galaxy types (SF, GV and QS), suggesting our conclusions remain unchanged.

\section{Conclusions}
We identify 456 gas-star kinematically misaligned galaxies from MPL-10 of MaNGA survey. We cross-match them with Chang's catalogue to obtain the global $M_*$ and SFR measurements, identifying 74 star-forming, 136 green-valley and 206 quiescent galaxies. We find that $\Delta\mathrm{PA}$ distribution has three local peaks around $0^{\circ}$, $90^{\circ}$ and $180^{\circ}$. The misaligned fraction peaks at $\log(M_*/M_{\odot})\sim10.5$ and declines to both low and high mass end. This fraction decreases monotonically with increasing SFR and sSFR. Through comparing the physical properties between misaligned galaxies and their control samples, we find that:

i) For SF and GV galaxies, the asymmetry of gas velocity fields is higher in misaligned galaxies than their control samples.

ii) The {\ion{H}{i}}-detection rate is lower in misaligned galaxies than their control samples in all the three types. The estimated molecular gas fraction based on the optical extinction is also lower for misaligned galaxies.

iii) Misaligned galaxies are smaller in effective radius for all the three types of galaxies. It is about $1.5-2~\mathrm{kpc}$ smaller for SF and GV misaligned galaxies than their controls, while it is $1~\mathrm{kpc}$ smaller for QS misaligned galaxies.

iv) Misaligned galaxies in all the three types show larger S\'{e}rsic index than their control samples, with the smallest difference in $n$ for QS galaxies. The median for SF misaligned galaxies is $n\sim3$, while the median value for control sample is less than 2. The GV misaligned galaxies (median $\sim4$) also show a more bulge dominated feature than controls (close to disc dominated, with median $\sim2$). Both QS misaligned galaxies and their controls display bulge dominated feature with the smallest difference in S\'{e}rsic index.

v) All the three types of misaligned galaxies have lower spin parameters $\lambda_{R_e}$ of stellar components compared to the control samples.

vi) Misaligned galaxies in all the three types have lower tidal strength parameter ($Q_{\mathrm{lss}}$) and similar trend is also found in neighbour count (Nneigh) with much less significance, suggesting that they tend to reside in more isolated environment.

Combining those observational results, we suggest that misaligned SF and GV galaxies form through gas accretion. The angular momentum experiences redistribution during the interaction between pre-existing and misaligned accreted gas, leading to gas inflow to the center which triggers central star formation. We propose three possible origins of the misaligned QS galaxies: (1) external gas accretion; (2) merger; (3) GV galaxies evolve into QS galaxies. 

\section*{Acknowledgements}
We thank the anonymous referee for the valuable comments which greatly improve the paper. Y.M.C. acknowledges support from the National Key R\&D Program of China (No. 2017YFA0402700), the National Natural Science Foundation of China (NSFC grants 11573013, 11733002, 11922302), the China Manned Space Project with NO.CMS-CSST-2021-A05. DB is partly support by grant RSCF 22-12-00080.

Funding for the Sloan Digital Sky 
Survey IV has been provided by the 
Alfred P. Sloan Foundation, the U.S. 
Department of Energy Office of 
Science, and the Participating 
Institutions. 

SDSS-IV acknowledges support and 
resources from the Center for High 
Performance Computing  at the 
University of Utah. The SDSS 
website is www.sdss.org.

SDSS-IV is managed by the 
Astrophysical Research Consortium 
for the Participating Institutions 
of the SDSS Collaboration including 
the Brazilian Participation Group, 
the Carnegie Institution for Science, 
Carnegie Mellon University, Center for 
Astrophysics | Harvard \& 
Smithsonian, the Chilean Participation 
Group, the French Participation Group, 
Instituto de Astrof\'isica de 
Canarias, The Johns Hopkins 
University, Kavli Institute for the 
Physics and Mathematics of the 
Universe (IPMU) / University of 
Tokyo, the Korean Participation Group, 
Lawrence Berkeley National Laboratory, 
Leibniz Institut f\"ur Astrophysik 
Potsdam (AIP),  Max-Planck-Institut 
f\"ur Astronomie (MPIA Heidelberg), 
Max-Planck-Institut f\"ur 
Astrophysik (MPA Garching), 
Max-Planck-Institut f\"ur 
Extraterrestrische Physik (MPE), 
National Astronomical Observatories of 
China, New Mexico State University, 
New York University, University of 
Notre Dame, Observat\'ario 
Nacional / MCTI, The Ohio State 
University, Pennsylvania State 
University, Shanghai 
Astronomical Observatory, United 
Kingdom Participation Group, 
Universidad Nacional Aut\'onoma 
de M\'exico, University of Arizona, 
University of Colorado Boulder, 
University of Oxford, University of 
Portsmouth, University of Utah, 
University of Virginia, University 
of Washington, University of 
Wisconsin, Vanderbilt University, 
and Yale University.

\section*{Data Availability}
The data underlying this article will be shared on reasonable request to the corresponding author.

\bibliographystyle{mnras}
\bibliography{reference.bib}



\label{lastpage}
\end{document}